\documentclass[a4paper,10pt]{article}    
\pdfoutput=1
\usepackage[OT1]{fontenc} 
\usepackage{multicol}
\usepackage{graphicx}   
\usepackage{bbold} 
\usepackage{mathtools}
\usepackage{amsmath} 
\numberwithin{equation}{section} 
\usepackage{amsfonts} 
\usepackage{amssymb}
\usepackage{amsbsy}
\usepackage{amsfonts}
\usepackage{physics}
\usepackage{bbold} 
\usepackage{slashed}   
\usepackage{color}
\usepackage[table]{xcolor} 
\usepackage{booktabs}
\usepackage{tabularx}
\usepackage{hyperref}
\usepackage{orcidlink}
\usepackage{bbding}  
\usepackage{tcolorbox}
\usepackage{ulem}

\definecolor{oucrimsonred}{rgb}{0.6, 0.0, 0.0}
\definecolor{DarkGray}{gray}{0.4}
\definecolor{lightgray}{gray}{0.9}
\definecolor{forestgreen}{rgb}{0.13,0.35,0.13}
\definecolor{ocre}{HTML}{F16723}

\usepackage[page,toc,titletoc,title]{appendix}
 
\numberwithin{equation}{section}
\numberwithin{table}{section}
\numberwithin{figure}{section}
\usepackage{bm}
\usepackage{cite}  
\usepackage{mathrsfs}
\usepackage{framed}
\usepackage[font=footnotesize,labelfont=bf]{caption}
\def\eq#1{{Eq.~(\ref{#1})}}
\def\eqs#1#2{{Eqs.~(\ref{#1})--(\ref{#2})}}

\def\abs#1{\left| #1\right|}

\def\Tr{\mbox{Tr}\,}

\colorlet{grayline}{gray!70}
\definecolor{blueline}{rgb}{0,0.27,0.55}
\definecolor{DarkGray}{gray}{0.4}
\definecolor{Gray}{gray}{0.6}
\definecolor{oucrimsonred}{rgb}{0.6, 0.0, 0.0}
\definecolor{persianblue}{rgb}{0.11, 0.22, 0.73}
\definecolor{forestgreen}{rgb}{0.13,0.35,0.13}
 \hypersetup{colorlinks, citecolor=forestgreen, linkcolor=forestgreen, urlcolor=forestgreen}
\newcommand{\be}{\begin{equation}}
\newcommand{\ee}{\end{equation}}
\newcommand{\bea}{\begin{eqnarray}}
\newcommand{\eea}{\end{eqnarray}}
\newcommand{\nn}{\nonumber}

\newcommand*\xbar[1]{%
  \hbox{\;%
    \vbox{%
      \hrule height 0.5pt 
      \kern0.5ex
      \hbox{%
        \kern-0.25em
        \ensuremath{#1}%
        \kern-0.07em
      }%
    }%
  }%
} 
\newcommand{\com}[1]{}
\newcommand{\gsim}{\lower.7ex\hbox{$\;\stackrel{\textstyle>}{\sim}\;$}}
\newcommand{\lsim}{\lower.7ex\hbox{$\;\stackrel{\textstyle<}{\sim}\;$}} 

\newcommand{\bc}{\begin{center}}
\newcommand{\ec}{\end{center}}
\newcommand{\K}{K^{*}(892)^0}

\newcommand{\red}[1]{\textcolor{red}{#1}}

\font\beeg=cmr17 scaled 1800
\newbox\ibox
\def\versal#1{\setbox\ibox=\hbox{{\beeg #1}~}%
	    \noindent\global\hangindent=\wd\ibox\global\hangafter-2%
	    \sc\smash{\llap {\lower 14pt \box\ibox}}}
\begin{document}

\thispagestyle{empty}

\vspace*{3cm}
\bc
{ \Large \color{oucrimsonred} \textbf{ 
Measuring CP violation using quantum state tomography}}
 \ec

\bc
\vspace*{1.5cm}
  {\bf M. Fabbrichesi$^{a\, \orcidlink{0000-0003-1937-3854}}$,}
{\bf   R. Floreanini$^{a\, \orcidlink{0000-0002-0424-2707}}$,}
{\bf E. Gabrielli$^{b,a,c\, \orcidlink{0000-0002-0637-5124}}$ and} 
 {\bf L. Marzola$^{c,d\, \orcidlink{0000-0003-2045-1100}}$}

\vspace{0.5cm}
{\small 
{\it  \color{DarkGray} (a)
INFN, Sezione di Trieste, Via Valerio 2, I-34127 Trieste, Italy}
\\[1mm]
  {\it \color{DarkGray}
    (b) Physics Department, University of Trieste, Strada Costiera 11,  I-34151 Trieste, Italy}
  \\[1mm]  
  {\it \color{DarkGray}
(c) Laboratory of High-Energy and Computational Physics, NICPB, R\"avala 10, 10143 Tallinn, Estonia}
 \\[1mm]
  {\it \color{DarkGray}
(d) Institute of Computer Science, University of Tartu, Narva mnt 18,  51009 Tartu, Estonia.}
}
\ec

 \vskip1cm
\bc
{\color{DarkGray}
\rule{0.7\textwidth}{0.5pt}}
\ec
\vskip3cm
\bc
{\bf ABSTRACT} 
\ec
\noindent We investigate direct CP violation in neutral meson decays by reconstructing the associated density matrices and measuring their difference using the trace distance. Our results cover neutral kaon decays into two scalar triplets of isospin space, specifically the pions, and decays of $B$- and $D$- mesons into two scalar octets of $SU(3)$ flavor space. We briefly discuss the quantum properties of these states, including entanglement, contextuality, and nonlocality. Additionally, we demonstrate a comparable approach for spin-1 final states by employing a density matrix describing states in the space of helicities. The significance of CP violation obtained through this method is consistently comparable, and often surpasses that obtained using only single, or combinations of, asymmetries.
\vspace*{1cm}

  \vskip 3cm
\bc 
{\color{DarkGray} 
\SquareShadowBottomRight
}
\ec

\newpage

\section{Introduction}
{\versal It is customary to express the violation}  of charge and parity (CP) in terms of asymmetries that quantify differences between the decays of a particle and of its CP-conjugated counterpart. This approach is natural because it aligns with the experimental methods used to measure CP violations, but other quantities may also be used. In particular, those expressed in terms of the parameters $\varepsilon$ and $\varepsilon^{\prime}$ are routinely used within kaon physics, where CP violation was first discovered and where it continues to be extensively studied. In addition to that, we presently have a growing amount of data on the $B$- and $D$-meson systems suitable for these analyses, as made manifest by the CP violation section in the Particle Data Group (PDG) review~\cite{ParticleDataGroup:2024cfk}.

\begin{figure}[h!]
\begin{center}
\includegraphics[width=3.5in]{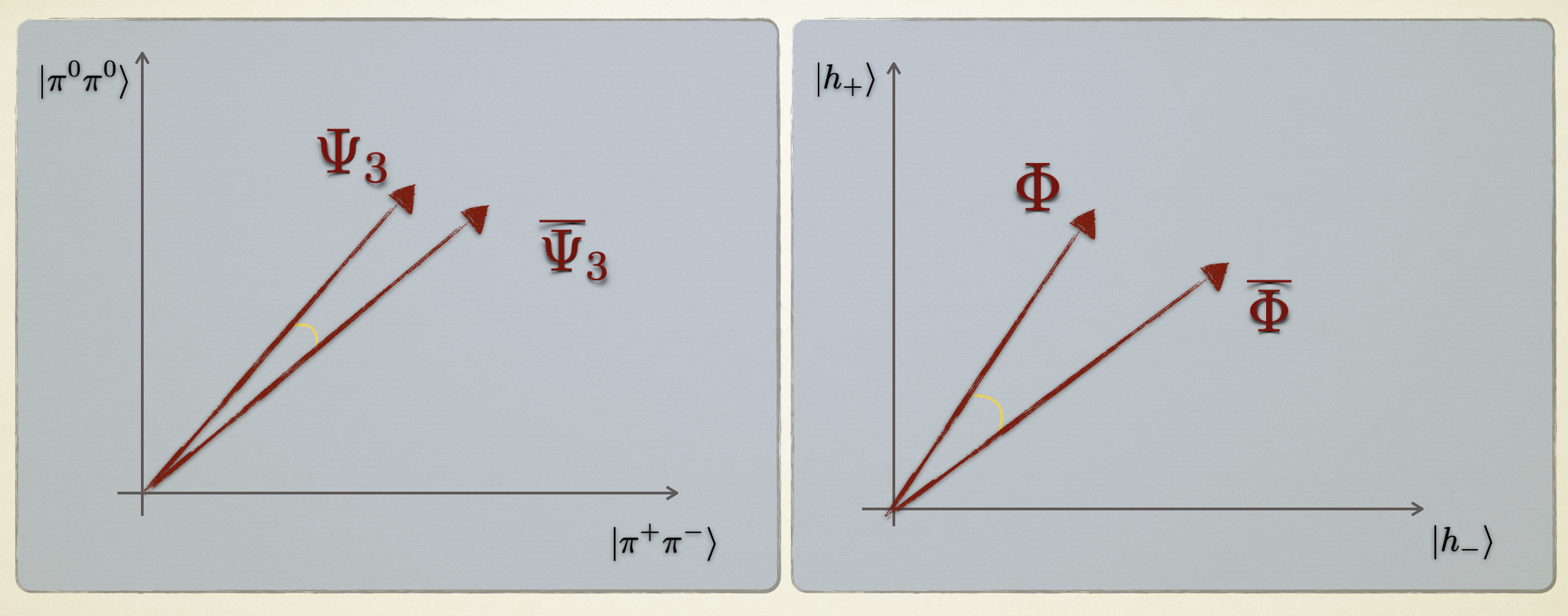}
 \caption{\footnotesize Examples of states in flavor space, $\Psi_3$ (left-hand side), and in helicity space, $\Phi$ (right-hand side), together with their CP-conjugated states $\overline \Psi_3$ and $\overline \Phi$. The axes are the bases of the two-dimensional Hilbert spaces: bipartite neutral states in isospin ($|\pi^i\pi^j\rangle$) and helicities ( ($|h_\pm\rangle$)), respectively. The difference between a state and its CP-conjugated partner is measured by the trace distance. 
\label{fig:draw}}
\end{center}
\end{figure}

In this work we put forward a distinct approach to the problem of quantifying CP violation in meson decays, one based on quantum state tomography. This procedure reconstructs the quantum state of a physical system by analyzing its properties within the relevant symmetry or parameter space, thereby providing a comprehensive description of the object under study. Fig.~\ref{fig:draw} provides a cartoon depiction of different possible quantum states in different spaces and their variation under a CP transformation. 

When studying the spin state of decaying particles or particle pairs, quantum state tomography utilizes the distribution of final state momenta to reconstruct spin correlations and polarizations, encoded in the related density matrix. For flavor space, instead, the required density matrices are obtained by considering the different amplitudes characterizing the various decays channels. For a thorough overview of quantum information concepts and techniques, the reader can choose from the numerous excellent books available~\cite{bruss2019quantum,Benatti:2010,Nielsen:2012yss,Wilde2017}; for the applications of this approach within high-energy physics, we recommend referring to~\cite{Barr:2024djo}. Additionally, for a more accessible introduction, please see~\cite{Fabbrichesi:2025aqp}.

As mentioned before, the final states generated in the decays of two CP-conjugated states, such as $B^0$ and $\overline{B}^0$, can be used to reconstruct the density matrices describing the quantum states of the original system through quantum state tomography. These matrices differ if the CP symmetry is not respected, and so the extent of CP violation can be measured by quantifying the difference between the density matrices. One possible way to do that is offered by the trace distance, which is a measure of the difference between the corresponding quantum states.

Given two density matrices $\rho$ and $\bar \rho$, the trace distance $\mathscr{D}^T(\rho,\, \bar \rho)$ is defined as~\cite{Nielsen:2012yss,Wilde2017}
\be
\mathscr{D}^T (\rho,\, \bar \rho) = \dfrac{1}{2} \, \Tr \sqrt{(\rho - \bar \rho)^\dag (\rho - \bar \rho)}\, . \label{eq:traceD}
\ee
If we take $\bar \rho$ to be the CP transformed of $\rho$, the trace distance can then be used as a measure of CP violation.\footnote{One could also use the fidelity distance as in
$
\mathscr{D}^F (\rho,\, \bar \rho) =  \sqrt{1-\left(\Tr \sqrt{\sqrt{\rho} \bar \rho \sqrt{ \rho}}\right)^2} \label{eq:traceF}
$
to measure the distance between the same matrices, but the two quantities yield the same value because the density matrices we will use describe pure states. } For pure states the expression simplifies to
\be
\mathscr{D}^T (|\Phi\rangle \langle \Phi|,|\overline \Phi\rangle \langle \overline\Phi|)= \sqrt{1-|\langle \Phi | \overline \Phi\rangle|^{2}}\,,
\ee
making it evident that $0\leq  \mathscr{D}^F (\rho,\, \bar \rho) \leq 1$. 
The lower bound is saturated by equal states, while the upper bound holds for orthogonal ones.    

The density matrices used to measure CP violation can be built by using states belonging to any space that describes the system under consideration. For instance, the tomographic procedure may use flavor, isospin or spin states, thereby testifying for the broad range of applicability of the method that we propose. The various channels through which the quantum systems under investigation may decay provide the bases of the Hilbert spaces describing for instance, the spin, flavor or isospin composition of the systems. The related quantum states are then defined by the weights of the basis elements, which can be experimentally determined by measuring the corresponding branching ratios or amplitudes. 

As mentioned before, the method we propose quantifies CP violation by measuring the difference between the quantum states of CP-conjugated systems. From this perspective, CP violation is then observed as a genuine property of the  particles under investigation rather than as an asymmetry inherent to a specific decay used to compare their properties. Consequently, we can  speak of measuring CP violation in the state defined by the decays of, for instance, the $B$-meson systems, rather than in a single, particular decay among the various channels in which it can decay. This bird’s-eye view is even more crucial because the origin of CP violation in the Standard Model can be traced back to a unique source, the  phase in the CKM matrix, which is shared across various decay channels.


\section{CP violation with isospin density matrices}

{\versal The decays of neutral kaons} into two pions were the first to reveal CP violation. The observation used measurements of the small but non-zero decay width of the CP-odd neutral kaon, $K_{L}$, into the CP-even two-pion final state. This system is the simplest one to treat with the methodology we propose and allows for a comparison with observables measuring direct CP violation, specifically the parameter $\varepsilon^\prime/\varepsilon$ that makes best use of the experimental data.

Let us begin by writing the state of two pions that originate from the decay of a kaon as
\be
\ket{\Psi_{3}} =\xi_1\,\Big[ |\psi_{\pi^+}(\vec{p}) \, \psi_{\pi^-}(-\vec{p}) \rangle + |\psi_{\pi^-} (\vec{p}) \, \psi_{\pi^+} (-\vec{p})\rangle\Big] +  \xi_2 \, |\psi_{\pi^0}(\vec{p}) \, \psi_{\pi^0}(-\vec{p})\rangle \,,
\label{state}
\ee
and the state of the pions produced in the decay of the CP-conjugated initial state:
\be
\ket{\overline \Psi_{3}} =\bar \xi_1\, \Big[ |\psi_{\pi^+}(\vec{p}) \, \psi_{\pi^-}(-\vec{p}) \rangle + |\psi_{\pi^-} (\vec{p}) \, \psi_{\pi^+} (-\vec{p})\rangle \Big]+  \bar\xi_2 \, |\psi_{\pi^0}(\vec{p}) \, \psi_{\pi^0}(-\vec{p})\rangle\, , \label{state1}
\ee
In the equations above, the momenta of the pions refer to the coordinate axes of the center-of-mass frame where the kaons decay at rest. 

Why are the two mesons in Eqs. \eqref{state} and \eqref{state1} in a coherent superposition instead of a mixture?
In the unbroken $SU(2)$ isospin limit, the isospin components of the pion state are indistinguishable. Consequently, they must be coherently summed and the final state produced is a superposition. This is the case of the ideal isospin-conserving strong interactions. Weak interactions and mass differences, being minor perturbations, cause only slight distortions in the eigenstates selected by strong interactions, while simultaneously generating a unitary evolution of a pure state into a pure state.

The density matrices describing the state of the two pions are $9\times 9$ matrices, each pion being in the adjoint representation of the group $SU(2)$. These density matrices are readily written in terms of the states in \eq{state} as
\be
 \rho_3 =    \frac{|\Psi_3 \rangle \langle \Psi_3|}{\langle \Psi_3|\Psi_3 \rangle }\quad  \text{and} \quad \bar\rho_3 =   \frac{|\overline  \Psi_3 \rangle\langle  \overline \Psi_3|}{\langle  \overline \Psi_3|\overline  \Psi_3 \rangle} \,.
\ee
More explicitly, on the basis where $\pi^+ = (1,0,0)^T$, $\pi^0 = (0,1,0)^T$ and $\pi^- = (0,0,1)^T$, we have
\be
\small \label{eq:rho3}
\rho_{3}=\frac{1}{2 |\xi _1|^2+|\xi _2|^2} \,
\left(
\begin{array}{ccccccccc}
 0 & 0 & 0 & 0 & 0 & 0 & 0 & 0 & 0 \\
 0 & 0 & 0 & 0 & 0 & 0 & 0 & 0 & 0 \\
 0 & 0 & |\xi _1|^2 & 0 & \xi _1 \xi _2^*& 0 & |\xi _1|^2&
   0 &0\\
 0 & 0 & 0 & 0 & 0 & 0 & 0 & 0 & 0 \\
 0 & 0 & \xi _2 \xi _1^*& 0 &  |\xi _2|^2  & 0& \xi _2 \xi _1^*& 0 &
   0 \\
 0 & 0 & 0 & 0 & 0 & 0 & 0 & 0 & 0 \\
 0 & 0 &  |\xi _1|^2& 0 & \xi _1 \xi _2^* & 0 & |\xi _1|^2& 0 &
   0 \\
 0 & 0 & 0 & 0 & 0 & 0 & 0 & 0 & 0 \\
 0 & 0 & 0 & 0 & 0 & 0 & 0 & 0 & 0 \\
\end{array}
\right)\,,
\ee 
and similarly for $\bar \rho_{3}$.

\subsection{Quantum-entangled pions redux}

The two-pion state resulting from the decay of the $K_S$ has already been discussed in~\cite{Fabbrichesi:2025zpw}, where it was found to exhibit quantum entanglement, contextuality, and Bell non-locality. The quantum entanglement is similar to that manifested by two particles with spin one produced in the decay of a scalar state, with the spin replaced by the isospin and the conservation of the angular momentum by the conservation of the charge. We briefly recap here the results and extend them to the decays of the $K_{L}$.

The quantum properties only depend on $\gamma$, the relative weight of the coefficients $\xi_{2}$ and $\xi_{1}$. In the case of the decays of $K_{S}$, 
\be
| \gamma|=\frac{|\langle\pi^{0}\pi^{0}| {\cal L}_{W}| K_{S}\rangle|}{|\langle\pi^{+}\pi^{-}| {\cal L}_{W}| K_{S}\rangle|}=0.9349\pm0.0011\,, \label{eq:gamma}
\ee 
in which ${\cal L}_W$ is the $\Delta S=1$ weak Lagrangian of the Standard Model. 
 
The reduced density matrix of one pion, obtained by tracing over the degrees of freedom of the other, describes an incoherent mixture of the three physical $\pi^+$, $\pi^0$ and $\pi^-$ states, with weights given by the elements on the diagonal. This reduced matrix determines whether the single pions produced in the decay features quantum contextuality, witnessed by the quantity~\cite{Fabbrichesi:2025ifv,Fabbrichesi:2025rsg} 
\be
 \mathbb{CNTXT}_9=
 \dfrac{8 \abs{\gamma}^2+19}{3 \abs{\gamma}^2+6}= 3.015\pm0.001 \label{cont9two}\ee
which, being significantly larger than 3, shows that the pions are in a contextual state.

As the pions are in a pure state, quantum entanglement can be quantified via the entropy of entanglement, 
the von Neumann entropy of the single pion reduced density matrix~\cite{Horodecki:2009zz}, equal to
\be
 \mathscr{E}[\rho]= \frac{1}{| \gamma |
   ^2+2} \Big[ | \gamma | ^2 \log \left(\frac{| \gamma | ^2+2}{|\gamma|^2}\right)  
   + 2  \log \left(| \gamma | ^2+2\right) \Big] = 1.097\pm0.001\, .
\ee
Finally, the quantity
\be
{\cal I}_3=\Tr[ \mathscr{B} \rho]= 2.894 \pm 0.001 \, ,
\ee
in which the matrix $\mathscr{B}$ is given in~\cite{Acin:2002zz}, can be used to witness Bell locality. For the two qutrits implemented by the pions, the inequality ${\cal I}_3\leq2$ is significantly violated.

The same holds true for the two-pion state created in the CP-violating decays of the $K_{L}$. It this case we find $\gamma^\prime$, the relative weight of the coefficients $\xi_{2}$ and $\xi_{1}$ in \eq{state1}, to be
\be
|\gamma^{\prime}|= \frac{|\langle\pi^{0}\pi^{0}| {\cal L}_{W}| K_{L}\rangle|}{|\langle\pi^{+}\pi^{-}| {\cal L}_{W}| K_{L}\rangle|}=0.930\pm0.002,
\ee
by using the widths given in \cite{ParticleDataGroup:2024cfk}. Accordingly,
\be
 \mathbb{CNTXT}_9=3.016 \pm0.001\, , \quad  \mathscr{E}[\rho]=1.096\pm0.001 \quad\text{and}\quad   {\cal I}_3= 2.895 \pm 0.001\,.
\ee
Therefore, the two-pion state originating in the decays of $K_{L}$ also exhibits comparable quantum entanglement, contextuality, and Bell non-locality. 

\subsection{CP symmetry properties} 

By comparing the density matrices for the decays of $K^0$ and $\overline K^0$, we can quantify the amount of direct CP violation encoded in the small but non-zero contribution from the decays of the $K_L$. 

We write the neutral kaon states as
\be
|K^{0}\rangle = \dfrac{1}{\sqrt{2}} \, \frac{\sqrt{1+\abs{\varepsilon}^2}}{1+\varepsilon} \Big( | K_{S}\rangle  + | K_{L}\rangle \Big) \quad \text{and} \quad |\overline K^{0}\rangle = \dfrac{1}{\sqrt{2}} \, \frac{\sqrt{1+\abs{\varepsilon}^2}}{1-\varepsilon} \Big( | K_{S}\rangle  - | K_{L}\rangle \Big) \, , \label{KK0}
\ee
in which $\varepsilon$ is the indirect CP violation parameter of the kaon system \cite{ParticleDataGroup:2024cfk}:
\be
\varepsilon = (2.280\pm 0.013)\times 10^{-3}\,e^{i \pi/4} . \label{eps}
\ee

The coefficients $\xi_{1,2}$ and $\bar \xi_{1,2}$ determining the two states generated in the decay of the $K^{0}$ and $\bar K^{0}$, respectively, can be derived by direct comparison:
\bea
\xi_1 &=&\left(1+\eta_{+-}\right)\frac{\sqrt{1+\abs{\varepsilon}^2}}{\sqrt{2}\, (1+\varepsilon)}\,\langle \pi^{+}\pi^{-} |{\cal L}_W| K_{S} \rangle\,, \quad \quad
\bar \xi_1=  \left(1-\eta_{+-}\right) \frac{\sqrt{1+\abs{\varepsilon}^2}}{\sqrt{2}\, (1-\varepsilon)}\, \langle \pi^{+}\pi^{-} |{\cal L}_W| K_{S} \rangle\,,\nn\\
\xi_2 &=& \left(1+\eta_{00}\right) \frac{\sqrt{1+\abs{\varepsilon}^2}}{\sqrt{2}\, (1+\varepsilon)}\, \langle \pi^{0}\pi^{0}| {\cal L}_W| K_{S} \rangle\,, \quad \quad
\bar \xi_2= \left(1-\eta_{00}\right) \frac{\sqrt{1+\abs{\varepsilon}^2}}{\sqrt{2}\, (1-\varepsilon)}\, \langle \pi^{0}\pi^{0}| {\cal L}_W| K_{S} \rangle\,.  \label{xi}
 \eea
The parameters  $\eta_{00}\equiv |\eta_{00}|e^{i\phi_{00}}$ and $\eta_{+-}\equiv |\eta_{+-}|e^{i\phi_{+-}}$ in \eq{xi} are defined as
 \be
 \eta_{00}= \frac{\langle\pi^{0}\pi^{0}| {\cal L}_{W}| K_{L}\rangle}{\langle \pi^{0}\pi^{0}| {\cal L}_W| K_{S} \rangle} \quad \text{and} \quad 
 \eta_{+-} =  \frac{\langle\pi^{+}\pi^{-}| {\cal L}_{W}| K_{L}\rangle}{\langle \pi^{+}\pi^{-}| {\cal L}_W| K_{S} \rangle}\,, \label{etas}
 \ee
and their experimental values are given in Table~\ref{tab:etas}. Physically, they match the CP asymmetries given in terms of branching ratios $\cal B$ as
\be
A_{CP} ( \pi^i\pi^j ) = \frac{{\cal B} (K^0\to \pi^i\pi^j )- {\cal B} (\bar K^0\to  \pi^i\pi^j)}{{\cal B} (K^0\to  \pi^i\pi^j)+ {\cal B} (\bar K^0\to  \pi^i\pi^j)}\, ,
\ee
as long as we neglect indirect CP violation, that is for $\varepsilon=0$;  the modulus of their ratio  is known with high precision~\cite{ParticleDataGroup:2024cfk}. In addition, we need
\be
\gamma = (0.9357 \pm0.0010) + i\, ( 0.0670\pm 0.0010)\label{eqgammaval}\, ,
\ee
as defined in \eq{eq:gamma}.
 
The phases in \eq{eps} and \eq{etas}  do not change sign under the CP operation that connects $K^{0}$ to $\bar K^{0}$, and the states differ only by the opposite sign of the $K_{L}$ term in \eq{KK0}. In this parametrization, the actual CP phase is encoded in the moduli of the utilized quantities, so that  trace distance and asymmetries are both proportional to the sine of that phase---and vanish as that phase vanishes.

\subsection{The trace distance at work} 

\begin{table}[h!]
\centering
\begin{tabular}{lc}
\toprule
 \textbf{parameter} & \textbf{value} \\
\midrule
$|\eta_{00}| $ & $(2.220\pm0.011)\times 10^{-3}$  \\
\rowcolor{lightgray} $|\eta_{+-}|$ & $ (2.232\pm0.011)\times 10^{-3}$ \\
$\phi_{00}$ & $ (43.52\pm0.05)^\circ  $ \\
\rowcolor{lightgray} $\phi_{+-}$ & $ (43.51\pm0.05)^\circ$ \\
$\left|\eta_{00} / \eta_{+-} \right|$ &$0.9950\pm 0.0007$ \\
\bottomrule
\end{tabular}
\caption{\label{tab:etas} Parameters entering the determination of the two-pion states in Eqs.~\eqref{state} and \eqref{state1}. Data from \cite{ParticleDataGroup:2024cfk}.}
\end{table}

CP violation in the kaon system induces a discrepancy between the density matrices $\rho_{3}$,  as defined in \eq{eq:rho3}, and $\bar \rho_{3}$.
Their difference can then be estimated by means of the trace distance in \eq{eq:traceD}. In fact, if the $K_L$ did not decay into two pions, CP symmetry would be restored, $\rho_3$ would match $\bar \rho_3$ and the trace distance $\mathscr{D}^T (\rho_3,\, \bar \rho_3)$ between them would vanish.

Some of the quantities entering this determination are correlated, as indicated by the comprehensive decay width fit performed by the PDG. We account for these correlation in the propagation of uncertainties, which we carry out using a Monte Carlo simulation. We use the double ratio $\left|\eta_{00} / \eta_{+-} \right|$  to input the difference in the moduli of  $\eta_{00}$ and $\eta_{+-}$, and use the values in Tab.~\ref{tab:etas} for the other inputs. This procedure yields the estimate
\be
\boxed{\mathscr{D}^T (\rho_3,\, \bar \rho_3)\Big|_{\text{CP}} =  \left(1.10 \pm 0.14\right) \times 10^{-5} \label{eq:D1}}
\ee
which implies a  non-vanishing value of the trace distance with a significance of about $7.2\sigma$.  As expected, if CP symmetry is preserved,  $\eta_{00}=\eta_{+-}$ and the trace distance vanishes. 

The significance of the trace distance in \eq{eq:D1} is comparable to that quoted for the conventional parametrization in terms of $\varepsilon^\prime/\varepsilon$ (approximately $7\sigma$). 
In considering this comparison, the reader should keep in mind that the high significance of the result can only be achieved in both the cases by including the double ratio $|\eta_{00}/\eta_{+-}|$, which is directly determined from the data. If only the values of $\eta_{00}$ and $\eta_{+-}$---whose uncertainty is at the percent level---were used, the significance of a non-vanishing $\varepsilon^\prime/\varepsilon$ ratio would reduce to about $1\sigma$, and similarly for the trace distance result.

\section{CP violation with $SU(3)$ flavor density matrices}

{\versal Measuring CP violation through the trace distance} can be applied to mesons related by the $SU(3)$ flavor symmetry, using the same approach employed for the kaon system. The $SU(3)$ scalar octet 
\be
\Pi= \begin{pmatrix} \eta/\sqrt{6} + \pi^0/\sqrt{2} & \pi^+/\sqrt{2} & K^+/\sqrt{2}\\
 \pi^-/\sqrt{2} &\eta/\sqrt{6} - \pi^0/\sqrt{2} & K^0/\sqrt{2}\\
  K^-/\sqrt{2}& \bar K^0/\sqrt{2} & - \sqrt{2/3} \,\eta
  \end{pmatrix}
\ee
gathers together states with different flavors: the neutral and charged pions, kaons, and the $\eta$-meson. In the internal flavor space of the group $SU(3)$, these particles can be considered as different components of a unique state---the octet seen by strong interactions.

The parameters
\be 
C_f = \frac{|A_f|^2 - |\xbar A_f|^2}{|A_f|^2 +|\xbar A_f|^2}  
\ee
are typically used to quantify the CP asymmetries in the decays of hadrons to a final state $f$ with amplitude $A_f$, and $\bar A_f$ for the decay of the CP-conjugated state into the same final state. Alternatively, the quantity
\be
A_{CP} (f) = \frac{{\cal B} (B\to f)- {\cal B} (\overline B\to \bar f)}{{\cal B} (B\to f)+ {\cal B} (\overline B\to \bar f )},
\ee
measures the asymmetries in the branching rates ${\cal B} (B\to f)$ for the meson $B$ (or $D$) to yield a final state $f$, with respect to the conjugated process (parent and daughters). The coefficient $C_f$ enters the time dependent asymmetry  in the term proportional to $\cos \Delta m_B\,t$, and is equivalent to the asymmetry $A_{CP}(f)$ if oscillations are neglected.

The density matrix $\rho_8$ and $\bar \rho_8$, which are  $64\times 64$ matrices, are impractical to reproduce here. They are constructed as 
\be
 \rho_8 =    \frac{|\Psi_8 \rangle \langle \Psi_8|}{\langle \Psi_8|\Psi_8 \rangle }\quad  \text{and} \quad \bar\rho_8 =   \frac{|\overline  \Psi_8 \rangle\langle  \overline \Psi_8|}{\langle  \overline \Psi_8|\overline  \Psi_8 \rangle} \,,
\ee
in which
the bipartite state $\Psi_8$ is defined as the coherent superposition 
\bea
|\Psi_8 \rangle &=& \xi_1 \,\Big[ |\psi_{\pi^+}(\vec{p}) \, \psi_{\pi^-}(-\vec{p}) \rangle + |\psi_{\pi^-} (\vec{p}) \, \psi_{\pi^+} (-\vec{p})\rangle \Big]+ 
 \xi_2 \, |\psi_{\pi^0}(\vec{p}) \, \psi_{\pi^0}(-\vec{p})\rangle +   \xi_3 \, |\psi_{\eta}(\vec{p}) \, \psi_{\eta}(-\vec{p}) \rangle   \label{eq:Psi}  \\
&+&  \xi_4 \,\Big[ |\psi_{K^+}(\vec{p}) \,\psi_{K^-}(-\vec{p}) \rangle + |\psi_{K^-} (\vec{p}) \, \psi_{K^+} (-\vec{p})\rangle \Big]+ 
 \xi_5  \,\Big[ |\psi_{K^0}(\vec{p}) \,\psi_{\bar K^0}(-\vec{p}) \rangle  + |\psi_{\bar K^0} (\vec{p}) \otimes \psi_{K^0} (-\vec{p}) \rangle \Big]\nn \\
 &+&  \xi_6 \,\Big[ |\psi_{K^+}(\vec{p}) \, \psi_{\pi^-}(-\vec{p}) \rangle + |\psi_{\pi^-} (\vec{p}) \, \psi_{K^+} (-\vec{p}) \rangle 
  \Big] +\xi_7 \,\Big[ |\psi_{K^-}(\vec{p}) \, \psi_{\pi^+}(-\vec{p}) \rangle + |\psi_{\pi^+} (\vec{p}) \, \psi_{K^-} (-\vec{p}) \rangle 
  \Big]  \nn \\
 &+&  \xi_8 \,\Big[ |\psi_{K^0}(\vec{p}) \,\psi_{\pi^0}(-\vec{p}) \rangle  +  |\psi_{\pi^0} (\vec{p}) \, \psi_{K^0} (-\vec{p})\rangle    \Big] +  \xi_9 \,\Big[ |\psi_{\bar K^0}(\vec{p}) \,\psi_{\pi^0}(-\vec{p}) \rangle  +  |\psi_{\pi^0} (\vec{p}) \, \psi_{\bar K^0} (-\vec{p})\rangle    \Big] \nn\\
  &+&  \xi_{10} \,\Big[ |\psi_{\pi^{0}}(\vec{p}) \,\psi_{\eta}(-\vec{p}) \rangle  +  |\psi_{\eta} (\vec{p}) \, \psi_{\pi^0} (-\vec{p})\rangle    \Big] +  \xi_{11} \,\Big[ |\psi_{\bar K^0}(\vec{p}) \,\psi_{\eta}(-\vec{p}) \rangle  +  |\psi_{\eta} (\vec{p}) \, \psi_{\bar K^0} (-\vec{p})\rangle    \Big]\, , \nn
 \eea
 and similarly with $\xi_i$ replaced by $\bar \xi_i$ for the CP conjugated state:
 \bea
|\overline \Psi_8 \rangle &=& \bar \xi_1 \,\Big[ |\psi_{\pi^+}(\vec{p}) \, \psi_{\pi^-}(-\vec{p}) \rangle + |\psi_{\pi^-} (\vec{p}) \, \psi_{\pi^+} (-\vec{p})\rangle \Big]+ 
\bar  \xi_2 \, |\psi_{\pi^0}(\vec{p}) \, \psi_{\pi^0}(-\vec{p})\rangle +   \bar \xi_3 \, |\psi_{\eta}(\vec{p}) \, \psi_{\eta}(-\vec{p}) \rangle   \label{eq:Psibar}  \\
&+& \bar  \xi_4 \,\Big[ |\psi_{K^+}(\vec{p}) \,\psi_{K^-}(-\vec{p}) \rangle + |\psi_{K^-} (\vec{p}) \, \psi_{K^+} (-\vec{p})\rangle \Big]+ 
\bar  \xi_5  \,\Big[ |\psi_{K^0}(\vec{p}) \,\psi_{\bar K^0}(-\vec{p}) \rangle  + |\psi_{\bar K^0} (\vec{p}) \otimes \psi_{K^0} (-\vec{p}) \rangle \Big]\nn \\
 &+&  \bar \xi_6 \,\Big[ |\psi_{K^+}(\vec{p}) \, \psi_{\pi^-}(-\vec{p}) \rangle + |\psi_{\pi^-} (\vec{p}) \, \psi_{K^+} (-\vec{p}) \rangle 
  \Big] +\bar \xi_7 \,\Big[ |\psi_{K^-}(\vec{p}) \, \psi_{\pi^+}(-\vec{p}) \rangle + |\psi_{\pi^+} (\vec{p}) \, \psi_{K^-} (-\vec{p}) \rangle 
  \Big]  \nn \\
 &+& \bar  \xi_8 \,\Big[ |\psi_{K^0}(\vec{p}) \,\psi_{\pi^0}(-\vec{p}) \rangle  +  |\psi_{\pi^0} (\vec{p}) \, \psi_{K^0} (-\vec{p})\rangle    \Big] + \bar  \xi_9 \,\Big[ |\psi_{\bar K^0}(\vec{p}) \,\psi_{\pi^0}(-\vec{p}) \rangle  +  |\psi_{\pi^0} (\vec{p}) \, \psi_{\bar K^0} (-\vec{p})\rangle    \Big] \nn\\
  &+& \bar  \xi_{10} \,\Big[ |\psi_{\pi^{0}}(\vec{p}) \,\psi_{\eta}(-\vec{p}) \rangle  +  |\psi_{\eta} (\vec{p}) \, \psi_{\pi^0} (-\vec{p})\rangle    \Big] + \bar  \xi_{11} \,\Big[ |\psi_{\bar K^0}(\vec{p}) \,\psi_{\eta}(-\vec{p}) \rangle  +  |\psi_{\eta} (\vec{p}) \, \psi_{\bar K^0} (-\vec{p})\rangle    \Big]\, . \nn
\eea

The two mesons  in \eqs{eq:Psi}{eq:Psibar}  are in a coherent superposition for the same reason already discussed in the case of the two pions produced in kaon decays. As before, the momenta in~\eqs{eq:Psi}{eq:Psibar} refer to the axes of the center-of-mass frame where the decaying meson is at rest. 
 
The weights $\xi_i$ in $|\Psi_8\rangle$ are the amplitudes for the corresponding final state, which we reconstruct from the decay widths $\Gamma_f$ as 
\be
\xi_f= \sqrt{\frac{\Gamma_f}{p(m_1,m_2)}}\,, \label{f}
\ee
after normalizing by the kinematical factors
\be
 p(m_1,m_2)=\sqrt{M_B^2 - (m_1+m_2)^2}\sqrt{M_B^2- (m_1-m_2)^2}\, .
\ee

\com{\red{$$ {\cal M} = T-i Pe^{i\delta} = \underbrace{T + P \sin\delta} + i P \cos \delta$$
$$ |{\cal M}|^2 = (T -i Pe^{i\delta} )(T+i Pe^{-i\delta} ) = T ^2+ 2 TP \sin\delta +P^2$$
$$\sqrt{|{\cal M}|^2} =\sqrt{T^2 + 2 TP \sin \delta + P^2} \simeq  \underbrace{T + P \sin \delta} + \frac{P^2}{2 T}$$
$$ e^{i \theta} \left( T + P \sin\delta + i P \cos \delta \right) = T + P \sin\delta $$
$$ {\cal M}_1 {\cal M}_2 = \underbrace{T_1 T_2 + (T_1P_2 + T_2 P_1) \sin \delta} + P_1 P_2 + i (P_1T_2-P_2T_1) \cos \delta$$
$$\sqrt{ |{\cal M}_1|^2} \sqrt{ |{\cal M}_2|^2} = \sqrt{T_1 T_2}  = \underbrace{T_1 T_2 + (T_1P_2 + T_2P_1) \sin \delta } + O(P_i^2) $$
}}

A comment is in order. The reconstruction of the amplitude in \eq{f} is incomplete because it is a real quantity, while the amplitude is generally complex. Nevertheless, this approximation is sufficient for our study. CP violation arises from the interference between the tree-level and penguin diagrams (the latter terms carrying the Cabibbo-Kobayashi-Maskawa (CKM) phase) that describe the meson decay. Our computation in \eq{f} of the amplitude neglects the contribution proportional to the cosine of the CKM phase of the penguin amplitude and higher-order terms in the penguin. Since these contributions are equal in both the density matrices $\rho_8$ and $\bar\rho_8$, excluding them does not affect the value obtained for the trace distance. It does however affect the full determination of the state. Although the state determined from \eq{f} is a good approximation, due to the penguin contribution being sub-leading, its complete determination is only possible with the full knowledge of the amplitudes. These amplitudes are known for the case of kaon decays in section 2 and helicity amplitudes in section 4, but not yet for the scalar octet.

In \eq{f}, the index $f$ runs overs the various bipartite systems $\pi\pi$, $KK$ and $\pi K$, created by the decay of the original mesons. Common factors have been dropped. The coefficients $\bar \xi_{i}$ in $|\overline \Psi_8\rangle$  are obtained from the widths and the asymmetries as
\be
\bar \xi_f=\xi_f \,\sqrt{\frac{1- A_{CP}(f)}{1+A_{CP}(f)}}\quad \text{or}\quad \bar \xi_f=\xi_f \,\sqrt{\frac{1- C_f}{1+C_f}} \, ,
\ee
depending on what asymmetry is known for the particular $f$ final state. CP violation can then be measured by analyzing the data for the decays of CP-conjugated systems. In the following  we consider the cases of the neutral  $B^{0}$ and $D^{0}$ mesons.

\subsection{Quantum-entangled scalar octet states}

The density matrices $\rho_{8}$ and $\bar \rho_{8}$ describing the quantum state formed by two members of the scalar meson octet can be studied to explore the quantum mechanical properties of this bipartite system. This is an intriguing extension of the approach used in the case of isospin space, as the space of flavor is defined by the group $SU(3)$, and the states are here qudits with eight components. In principle, quantum information quantities such as entanglement entropy and nonlocality can be measured using witnesses.

In the case of the $B^{0}$ meson, the entropies of entanglement~\cite{Barr:2024djo} of the flavor states are
\be
\mathscr{E}[\rho_{8}] = 1.670\pm0.008\,, \qquad \mathscr{E}[\bar{\rho}_{8}] = 1.667\pm0.008\,.
\ee
In the case of the $D^{0}$ meson, instead, the entropies of entanglement are given by
\be
\mathscr{E}[\rho_{8}] = 1.321\pm0.004\,,\quad \mathscr{E}[\bar{\rho}_{8}] = 1.322\pm0.004\,.
\ee
Therefore, both mesons yield states that are highly entangled in $SU(3)$ flavor space, and the presence of entanglement is evident with a significance well above the $5\sigma$ mark. 

While these numerical results are valid only within our approximation of neglecting part of the penguin amplitude, the values for the $\rho_{8}$ are close to the correct ones due to the smallness of the penguin contribution itself. Furthermore, the similarity in the values of the entanglement between CP conjugated decays arises from their differing only by the small contribution of the interference between the leading tree diagram and the penguin term, which is proportional to the sine of the CKM phase.

Unfortunately, there is no straightforward way to compute contextuality and Bell nonlocality witnesses for qudits of dimension 8.

\subsection{The trace distance at work} 

\begin{table}[h!]
\centering
\begin{tabular}{lcc}
\toprule
 \textbf{Decay} & \textbf{BR ($\times 10^{-5}$)} & \textbf{Asymmetry} \\
\midrule
$B^0 \to \pi^0 \pi^0$ & $0.155 \pm 0.017$ & $C_{\pi^0 \pi^0} = -0.25 \pm 0.20$ \\
\rowcolor{lightgray} $B^0 \to \pi^+ \pi^-$ & $0.537 \pm 0.020$ & $C_{\pi^+ \pi^-} = -0.314 \pm 0.030^{*}$ \\
$B^0 \to K^0_S K^0_S$ & $0.121 \pm 0.016$ & $C_{K^0_S K^0_S} = 0.0 \pm 0.4$ \\
\rowcolor{lightgray}$B^0 \to K^+ K^-$ & $0.0078 \pm 0.0015$ & -- \\
$B^0 \to \pi^0 K^0_S$ & $1.01 \pm 0.04$ & $C_{\pi^0 K^0_S} = 0.00 \pm 0.08$ \\
\rowcolor{lightgray}$B^0 \to \pi^- K^+$ & $2.00 \pm 0.04$ & $A_{CP}(K^- \pi^+) = -0.0831 \pm 0.0031^{*}$ \\
$B^0 \to \pi^0 \eta$ & $0.041 \pm 0.017$ & -- \\
\rowcolor{lightgray}$B^0 \to K^0_S \eta$ & $0.123 \pm 0.027$ & -- \\
$B^0 \to \eta \eta$ & $< 0.1$ & -- \\
\bottomrule
\end{tabular}
\caption{\label{tab:data2} Branching ratios (BR) and asymmetries for the decays of $B^0$ into two members of the scalar octet. Quantities with an asterisk allow for the observation of CP violation. Data from \cite{ParticleDataGroup:2024cfk}.}
\end{table}

For the $B$-meson decays, we include in the density matrices $ \bar \rho_{8}$  all the entries computed from the values given in Table~\ref{tab:data2}, which summarizes the present experimental results available. The uncertainties are propagated after including the correlations reported with the PDG fit. Some of the uncertainties are large enough that the linear approximation fails to describe the propagation and the final distribution of the uncertainty affecting the trace distance is rather skewed.  Since to estimate the significance of the proposed method we are only interest in the values between zero and the mode of the distribution, we model this region of the distribution after a Gaussian distribution centered on the mode. The significance of the obtained trace distance value can then be quantified by simply measuring the distance of the peak from the origin in terms of standard deviations. 

We thus obtain
\be
\boxed{\mathscr{D}^T (\rho_8,\, \bar \rho_8)\Big|_{\text{CP}} =  0.11\pm 0.02 \label{eq:D3}}
\ee
which implies that the obtained value of the trace distance has a significance of about $5.5\sigma$. 

The reason why the above significance is smaller than that of the asymmetry measured in the $K^{-}\pi^{+}$ decay channel is simple to understand. The value of this asymmetry is very small and is washed out when combined with those of other asymmetries that are much larger and dominate the distance between the CP-conjugated states. Had we only kept the latter asymmetry we would have obtained a comparable significance.

The same procedure can be followed for the decays of the $D$-meson. We include in the density matrices all the entries computed from the asymmetries---even though none of them is significantly different from zero. 

\begin{table}[h!]
\centering
\begin{tabular}{lcc}
\toprule
\textbf{Decay} & \textbf{BR ($\times 10^{-2}$)} & \textbf{Asymmetry} \\
\midrule
$D^0 \to \pi^0 \pi^0$ & $0.0826 \pm 0.0025$ & $A_{CP}(\pi^0 \pi^0) = 0.000 \pm 0.006$ \\
\rowcolor{lightgray}
$D^0 \to \pi^+ \pi^-$ & $0.1453 \pm 0.0024$ & $A_{CP}(\pi^+ \pi^-) = 0.0013 \pm 0.0014$ \\
$D^0 \to K^0_S K^0_S$ & $0.0141 \pm 0.0005$ & $A_{CP}(K^0_S K^0_S) = -0.011 \pm 0.019$ \\
\rowcolor{lightgray}
$D^0 \to K^+ K^-$ & $0.408 \pm 0.006$ & $A_{CP}(K^+ K^-) = (4 \pm 5) \times 10^{-4}$ \\
$D^0 \to \pi^0 K^0_S$ & $1.240 \pm 0.022$ & $A_{CP}(\pi^0 K^0_S) = -0.0020 \pm 0.0017$ \\
\rowcolor{lightgray}
$D^0 \to \pi^+ K^-$ & $3.945 \pm 0.03$ & $A_{CP}(\pi^+ K^-) = 0.002 \pm 0.005$ \\
$D^0 \to \pi^- K^+$ & $0.0150 \pm 0.0007$ & $A_{CP}(\pi^- K^+) = -0.009 \pm 0.014$ \\
\rowcolor{lightgray}
$D^0 \to K^0_S \eta$ & $0.434 \pm 0.016$ & $A_{CP}(K^0_S \eta) = 0.005 \pm 0.005$ \\
$D^0 \to \eta \eta$ & $0.211 \pm 0.019$ & -- \\
\bottomrule
\end{tabular}
\caption{\label{tab:data3} Branching ratios (BR) and asymmetries for the decays of $D^0$ into two members of the scalar octet. Data from \cite{ParticleDataGroup:2024cfk}.}
\end{table}

Using the values given in Table~\ref{tab:data3}, and again propagating the uncertainties after including the correlation indicated by the PDG fit, we find a distribution with  skewness of about 0.7. As before, we  approximate  the side toward the origin as a half-Gaussian distribution centered on the mode of the actual distribution, thus obtaining
\be
\boxed{\mathscr{D}^T (\rho_8,\, \bar \rho_8)\Big|_{\text{CP}} =  \left(3.31\pm 1.02 \right)\times 10^{-3}\,, \label{eq:D4}}
\ee
which implies that the trace distance does not vanish with at a significance of about $3.3\sigma$.

Using the trace distance, we have confirmed the presence of CP violation in the $D$-meson system. Prior to this, CP violation was only observed, with a significance better than $3\sigma$, in the difference between the CP asymmetries in the  decays of the $D^0$ into $K^+K^-$ and $\pi^+\pi^-$, as reported in~\cite{LHCb:2019hro}.

\section{CP violation with helicity density matrices}

{\versal As a second class of examples} which show how the trace distance can be used to study CP violation, we analyze the same decay process of the $B$ meson paying attention to the spin state of the decay products.  The approach used in the previous sections can be immediately applied to this scenario as well.

Let us consider the case of the polarization amplitudes to study the decay of a scalar state into two spin-1 particles. The bipartite quantum state formed by the latter is given by
\be
|\Phi \rangle=\frac{1}{\sqrt{|h_+|^2+h_0^2+|h_-|^2}} \Big[ h_+\, |\psi_{+}(\vec{p}) \, \psi_{-}(-\vec{p}) \rangle + h_0\, |\psi_{0} (\vec{p}) \, \psi_{0} (-\vec{p})\rangle +  h_-\, |\psi_{-}(\vec{p}) \, \psi_{+}(-\vec{p})\rangle\Big] \label{phi}\, ,
\ee
in which $\psi_{\pm, 0}$ represent the wave-function for the corresponding helicity and the momenta refer to the axes of the center-of-mass frame, where the $B$ meson decays at rest; of the three coefficients $h_i$, $h_0$ can be taken to be real,
while $h_\pm \equiv|h_\pm|\, e^{i\phi_\pm}$.

The corresponding density operator is represented by a $9\times 9$ matrix, $\rho=\dyad{\Phi}{\Phi}$, explicitly given by
\be\small
\rho=\dfrac{1}{|{\cal M}|^2} \begin{pmatrix}
 0 & 0 & 0 & 0 & 0 & 0 & 0 & 0 & 0 \\
 0 & 0 & 0 & 0 & 0 & 0 & 0 & 0 & 0 \\ 
 0 & 0 & |h_+|^2 & 0 & h_0 |h_+| e^{i \phi_+} &
   0 & |h_+| |h_-| e^{-i (\phi_- - \phi_+)} & 0 & 0 \\
 0 & 0 & 0 & 0 & 0 & 0 & 0 & 0 & 0 \\
 0 & 0 & h_0 |h_+| e^{-i \phi_+} & 0 & h_0^2&
   0 & h_0 |h_-| e^{-i \phi_-}& 0 & 0 \\
 0 & 0 & 0 & 0 & 0 & 0 & 0 & 0 & 0 \\
 0 & 0 &  |h_+| |h_-| e^{i (\phi_- - \phi_+)}  & 0 & h_0 |h_-| e^{i \phi_-}
   & 0 & |h_-|^2 & 0 & 0 \\
 0 & 0 & 0 & 0 & 0 & 0 & 0 & 0 & 0 \\
 0 & 0 & 0 & 0 & 0 & 0 & 0 & 0 & 0 \\
\end{pmatrix},
\ee
in which $h_\pm = (A_\parallel \pm A_\perp)/\sqrt{2}$ and $h_0 = A_0$ and $|{\cal M}|^2=|h_+|^2 +h_0^2 +|h_-|^2 $. The quantities $A_\parallel$, $A_\perp$ and $A_0$ are the polarization amplitudes given by experiments; we take the latter amplitude to be real. The density matrix describing the CP-conjugated state then is
\be\small
\bar \rho= \dfrac{1}{|{\cal M}|^2} \begin{pmatrix}
 0 & 0 & 0 & 0 & 0 & 0 & 0 & 0 & 0 \\
 0 & 0 & 0 & 0 & 0 & 0 & 0 & 0 & 0 \\
 0 & 0 & |h_-|^2 & 0 & h_0 |h_-| e^{-i \phi_-} &
   0 & |h_-| |h_+| e^{i (\phi_+ - \phi_-)} & 0 & 0 \\
 0 & 0 & 0 & 0 & 0 & 0 & 0 & 0 & 0 \\
 0 & 0 & h_0 |h_-| e^{i \phi_-} & 0 & h_0^2&
   0 & h_0 |h_+| e^{i \phi_+}& 0 & 0 \\
 0 & 0 & 0 & 0 & 0 & 0 & 0 & 0 & 0 \\
 0 & 0 &  |h_-| |h_+| e^{-i (\phi_+ - \phi_-)}  & 0 & h_0 |h_+| e^{-i \phi_+}
   & 0 & |h_+|^2 & 0 & 0 \\
 0 & 0 & 0 & 0 & 0 & 0 & 0 & 0 & 0 \\
 0 & 0 & 0 & 0 & 0 & 0 & 0 & 0 & 0 \\
\end{pmatrix}.
\ee

The trace distance $\mathscr{D}^T (\rho,\, \bar \rho)$ vanishes when
\be
|h_-|^2 =|h_+|^2 \quad \text{and} \quad \phi_- +   \phi_+ = 0 \, , \label{cond}
\ee
which, for the polarization amplitudes, gives
\be
|A_\lambda|=|\bar A_{-\lambda}| \quad \text{and} \quad \arg A_\lambda = \arg \bar A_{-\lambda}\, ,
\ee
implying that CP is conserved.
\begin{table}[h!]
\centering
\begin{tabular}{lcc}
\toprule
\textbf{Quantity} & \textbf{$B^0 \to J/\psi K^*$} & \textbf{$\overline{B}^0 \to J/\psi K^*$} \\
\midrule
$|A_\parallel|^2$ & $0.230 \pm 0.005$ & $0.225 \pm 0.005$ \\
\rowcolor{lightgray}
$|A_\perp|^2$ & $0.194 \pm 0.005$ & $0.207 \pm 0.005$ \\
$\delta_\parallel$ [rad] & $-2.94 \pm 0.03$ & $-2.92 \pm 0.03$ \\
\rowcolor{lightgray}
$\delta_\perp$ [rad] & $2.94 \pm 0.02$ & $2.96 \pm 0.02$ \\
\bottomrule
\end{tabular}
\caption{\label{tab:dataA} Results from the fit of the decays $B^0 \to J/\psi K^*$ and $\overline{B}^0 \to J/\psi K^*$ 
for the different polarization amplitudes, for modulus and relative phase, and
$|A_0|^2 = 1 - |A_\parallel|^2 - |A_\perp|^2$. 
From~\cite{LHCb:2013vga}.}
\end{table}

\begin{table}[h!]
\centering
\begin{tabular}{lc}
\toprule
\textbf{Quantity} & \textbf{$B_s^0 \to J/\psi\,\phi$} \\
\midrule
$|A_\parallel|^2$ & $0.2220 \pm 0.0017 \pm 0.0021$ \\
\rowcolor{lightgray}
$|A_0|^2$ & $0.5152 \pm 0.0012 \pm 0.0034$ \\
$\delta_\parallel$ [rad] & $3.36 \pm 0.05 \pm 0.09$ \\
\rowcolor{lightgray}
$\delta_\perp$ [rad] & $3.22 \pm 0.10 \pm 0.05$ \\
$\phi_s$ [rad] & $-0.087 \pm 0.036 \pm 0.021$ \\
\bottomrule
\end{tabular}
\caption{\label{tab:dataB} Results from the fit of the decays $B_s^0 \to J/\psi\,\phi$ for the different polarization amplitudes, 
for modulus and relative phase, and $|A_0|^2 = 1 - |A_\parallel|^2 - |A_\perp|^2$, with $\phi_s$ the CP-violating phase.
We use the combination of data from the 13~TeV run and the 7 and 8~TeV runs, solution (a).
See the analysis in~\cite{ATLAS:2020lbz} for the error covariance matrix. 
Errors are statistical and systematic.}
\end{table}

\subsection{Quantum-entangled spins} 

Experiments have reported data for two processes, $B^{0}\to J/\psi \K$ and  $B^{0}_s\to J/\psi \phi$, which we both analyze. 

The bipartite system formed by the two spin-1 final state mesons exhibits several quantum properties of interest that can be witnessed by means of the same operators utilized in section 2 for the two-pion system. Entanglement, contextuality and nonlocality have been previously discussed in \cite{Fabbrichesi:2023idl} and \cite{Fabbrichesi:2025rsg} within the present context; we briefly review these results and compare them with those produced by the CP-conjugated state.

The $J/\psi$ meson in the decay of the $B^0$ shows contextuality as
\be
\mathbb{CNTXT}_9=3.18\pm0.01 
\ee
is significantly larger than 3, and so does the $\K$. The bipartite system of the $J/\psi$ and the $\K$ exhibits also entanglement and nonlocality: 
\be
\mathscr{E}[\rho]=  0.756\pm0.009\, ,
 \quad \text{and} \quad {\cal I}_3=\Tr[ \mathscr{B} \rho]= 2.549 \pm 0.018\, .
  \ee
Similar results hold for the $J/\psi$ and $\phi$ in the decay of the $B^0_s$. 

The decay of the CP-conjugated state exhibit similar quantum properties. For instance, for the $J/\psi$ has a contextuality 
\be
\mathbb{CNTXT}_9=3.18\pm0.01\, , 
\ee
and, for the bipartite system made of the $J/\psi$ and $\K$ mesons, we observe the following results
\be
\mathscr{E}[\rho]= 0.757 \pm 0.010
\quad \text{and} \quad {\cal I}_3=\Tr[ \mathscr{B} \rho]= 2.548 \pm 0.018.
\ee
As in the case of the two-pion system, we do not find much difference between the quantum properties of the CP-conjugated systems.

\subsection{The trace distance at work}

Considering first the decay  $B^{0}\to J/\psi \K$, the experimental collaborations have provided results pertaining to the helicity amplitudes involved in the decay of the $B^{0}$ and the $\overline B^{0}$ mesons. The  measurements are correlated and the correlations can be found in~\cite{ATLAS:2020lbz}.
Using the values given in Table~\ref{tab:dataA}, we propagate the correlated uncertainties through a Monte Carlo simulation. This time, the size of the the various helicity components and their asymmetries are comparable and the final distribution is very close to be Gaussian, so that we can quote the mean and standard deviation: 
\be
\boxed{\mathscr{D}^T (\rho,\, \bar \rho)\Big|_{\text{CP}} = 0.028 \pm 0.006\, .}
\ee
The resulting significance is $4.7\sigma$, better than the $2\sigma$ obtainable with the helicity amplitudes themselves.

For the $B^{0}_s\to J/\psi \phi$ decay, the experimental collaborations have provided the helicity amplitudes for the $B^{0}_s$ decay, together with a determination of the the weak, CP-violating phase. Accordingly, the density matrix $ \bar \rho$ is obtained from $\rho$ by changing the sign of this weak phase. By using the values given in Table~\ref{tab:dataB}, we find, after propagating the correlated uncertainties,
\be
\boxed{\mathscr{D}^T (\rho,\, \bar \rho)\Big|_{\text{CP}} = 0.087 \pm 0.040\, ,}
\ee
which is non-vanishing with a significance of about $2\sigma$ (comparable to the significance of a non-vanishing $\phi_s$ phase).

\section{The trace distance vs.\ a $\chi^{2}$ test} 

{\versal In this section we address the question} of whether, and under what conditions, the trace distance outperforms a simple $\chi^{2}$ test. 

The combination of independent observations using a $\chi^{2}$ test does not always enhance the test’s power. Starting with a single measurement, characterized by an uncertainty of $1\sigma$ and resulting in a $\chi^{2}=1$, we anticipate that 68\% of the events fall within the $1\sigma$ interval. However, adding more measurements with the same relative uncertainty reduces the percentage of events because the increased value of $\chi^{2}$ is offset by the change in the $\chi^{2}$-distribution due to the additional degrees of freedom. For instance, three measurements that all deviate by $1\sigma$ from the corresponding best fit yield a $\chi^{2}=3$, which results in only 61\% of the events being included in the corresponding interval. The use of trace density is more effective in this scenario because it avoids the dilution caused by the additional degrees of freedom while still accounting for their impact in the density matrices. This is demonstrated in the case of neutral $D$-meson decays in section 3 and the polarization states in section 4.

The advantage provided by the trace distance is lost when the measurement pool consists of one, or a few, highly precise measurements. In such cases, the uncertainty associated with the trace distance tends to align with that of the less precise measurement, which ultimately dominates the uncertainties impacting the tomographic procedure. As previously mentioned, when a more uncertain measurement is incorporated into the $\chi^{2}$ test, the increase in the $\chi^{2}$ value is usually insufficient to offset the reduction in the number of degrees of freedom. Nevertheless, the significance of the test remains unaffected as long as the p-value obtained using the $\chi^{2}$ distribution approaches the maximum value of 1---a value for which the cumulative distribution function reaches its saturation point. This concept is exemplified by the case of the neutral $B$-meson decays discussed in section 3.

\section{Outlook} 

{\versal Quantum state tomography} offers a novel way to compare the quantum states of CP-conjugated systems based on the analysis of their decay products. A difference in the reconstructed quantum states would signal the presence of CP violation, and the latter could then be quantified by means of standard quantum information methods routinely used to compare quantum system. In this paper, after reconstructing the quantum states of a series of CP-conjugated physical systems, we explored the possible presence of CP-violation by using the trace distance as a diversity measure. 

We have argued that the proposed method offers a novel way to describe the amount of CP violation in a system, and that it demonstrates a superior sensitivity if compared to conventional measures based on asymmetries. It is most effective when data on multiple decay channels are available, and the corresponding CP-violating asymmetries are comparable in magnitude. The tomographic procedure is easily implemented for two-body decays because the relationship between decay widths and amplitudes is, in this case, straightforward. The extension to three-body decays presents more significant challenges.

Future studies on CP violation in kaon, $B$-, and $D$-meson systems, as well as further analyses of the polarization amplitudes of CP-violating decays, can be significantly improved by incorporating this novel method alongside the existing ones.

\small
\bc
{\bf Aknowledgements} 
\ec
LM is supported by the Estonian Research Council under the RVTT3, TK202 and PRG1884 grants.

\small
\bibliographystyle{JHEP}   
\bibliography{cp.bib}

\end{document}